\documentclass[conference]{IEEEtran}
\IEEEoverridecommandlockouts
\usepackage{import}
\usepackage{enumitem}
\usepackage{listings}
\usepackage{soul,color}
\usepackage{subcaption}
\usepackage{hyperref}
\hypersetup{
  hidelinks,
  colorlinks=true,
  allcolors=black,
  pdfstartview=Fit,
  breaklinks=true
}
\usepackage{multicol}
\usepackage{array, booktabs}
\usepackage{cite}
\usepackage{amsmath,amssymb,amsfonts}
\usepackage{algorithmic}
\usepackage{graphicx}
\usepackage{textcomp}
\usepackage{xcolor}
\usepackage{flushend}
\usepackage{tcolorbox}
\PassOptionsToPackage{table}{xcolor}

\def\BibTeX{{\rm B\kern-.05em{\sc i\kern-.025em b}\kern-.08em
    T\kern-.1667em\lower.7ex\hbox{E}\kern-.125emX}}
\begin{document}

% \title{Conference Paper Title*\\
% {\footnotesize \textsuperscript{*}Note: Sub-titles are not captured in Xplore and
% should not be used}
% \thanks{Identify applicable funding agency here. If none, delete this.}
% }

\title{Quantum Software Analytics: \\ Opportunities and Challenges}

\author{\IEEEauthorblockN{Thong Hoang\IEEEauthorrefmark{1}, Hoa Khanh Dam\IEEEauthorrefmark{2}, Tingting Bi\IEEEauthorrefmark{1}, Qinghua Lu\IEEEauthorrefmark{1}\IEEEauthorrefmark{3}, \\ Zhenchang Xing\IEEEauthorrefmark{1}\IEEEauthorrefmark{4}, Liming Zhu\IEEEauthorrefmark{1}\IEEEauthorrefmark{3}, Lam Duc Nguyen\IEEEauthorrefmark{1}, Shiping Chen\IEEEauthorrefmark{1}}

\IEEEauthorblockA{\IEEEauthorrefmark{1}\textit{CSIRO’s Data61}, \IEEEauthorrefmark{2}\textit{University of Wollongong}, \IEEEauthorrefmark{3}\textit{University of New South Wales}, \IEEEauthorrefmark{4}\textit{Australian National University}\\
\{james.hoang, tingting.bi, qinghua.lu, zhenchang.xing, liming.zhu, lam.nguyen, shiping.chen\}@data61.csiro.au}

\IEEEauthorblockA{\{hoa\}@uow.edu.au}
}

% \author{\IEEEauthorblockN{1\textsuperscript{st} Given Name Surname}
% \IEEEauthorblockA{\textit{dept. name of organization (of Aff.)} \\
% \textit{name of organization (of Aff.)}\\
% City, Country \\
% email address or ORCID}
% \and
% \IEEEauthorblockN{2\textsuperscript{nd} Given Name Surname}
% \IEEEauthorblockA{\textit{dept. name of organization (of Aff.)} \\
% \textit{name of organization (of Aff.)}\\
% City, Country \\
% email address or ORCID}
% \and
% \IEEEauthorblockN{3\textsuperscript{rd} Given Name Surname}
% \IEEEauthorblockA{\textit{dept. name of organization (of Aff.)} \\
% \textit{name of organization (of Aff.)}\\
% City, Country \\
% email address or ORCID}
% \and
% \IEEEauthorblockN{4\textsuperscript{th} Given Name Surname}
% \IEEEauthorblockA{\textit{dept. name of organization (of Aff.)} \\
% \textit{name of organization (of Aff.)}\\
% City, Country \\
% email address or ORCID}
% \and
% \IEEEauthorblockN{5\textsuperscript{th} Given Name Surname}
% \IEEEauthorblockA{\textit{dept. name of organization (of Aff.)} \\
% \textit{name of organization (of Aff.)}\\
% City, Country \\
% email address or ORCID}
% \and
% \IEEEauthorblockN{6\textsuperscript{th} Given Name Surname}
% \IEEEauthorblockA{\textit{dept. name of organization (of Aff.)} \\
% \textit{name of organization (of Aff.)}\\
% City, Country \\
% email address or ORCID}
% }

\maketitle

\begin{abstract}
Quantum computing systems depend on the principles of quantum mechanics to perform multiple challenging tasks more efficiently than their classical counterparts. In classical software engineering, the software life cycle is used to document and structure the processes of design, implementation, and maintenance of software applications. It helps stakeholders understand how to build an application. In this paper, we summarize a set of software analytics topics and techniques in the development life cycle that can be leveraged and integrated into quantum software application development. The results of this work can assist researchers and practitioners in better understanding the quantum-specific emerging development activities, challenges, and opportunities in the next generation of quantum software.
\end{abstract}


\section{Introduction}
\label{sec:intro}

\textit{Quantum computing} (QC) has emerged as the future for solving many problems more efficiently. For example, QC is used to simulate complex biochemical systems~\cite{cheng2020application}, reduce the training time of machine learning models~\cite{biamonte2017quantum}, and create encryption methods for preventing cybersecurity threats~\cite{mosca2018cybersecurity}. Unlike classical computing, where the information is encoded as bits and each bit is assigned either 0 or 1, QC encodes the information as a list of quantum bit (qubit). Each qubit is a linear combination of two qubit states, such as $|0\rangle$ and $|1\rangle$. In recent years, the development of quantum computing systems has attracted significant interests from both research and industry communities ~\cite{bohm2012quantum, montanaro2016quantum, chiribella2008quantum}. Cloud-based quantum computing platforms have emerged to enable developers to create quantum software applications. For example, Google has created a Quantum Virtual Machine\footnote{\url{https://quantumai.google/quantum-virtual-machine}} to emulate the results of quantum computers. IBM has offered a cloud quantum platform, namely IBM Quantum,\footnote{\url{https://quantum-computing.ibm.com/}} to help developers run their programs on quantum systems.
% As QC employs the principles of quantum mechanics to process data, it has the potential to efficiently outperform classical computing in solving problems across various domains, such as biochemistry, machine learning, cybersecurity, etc. 
%\hoa{Cite some industry quantum platform like Google, etc.}

% The main objective of QSE is to create quantum software applications throughout a quantum software life cycle~\cite{zhao2020quantum}. 

There has been growth in the number of quantum-driven software systems in recent years. Hence, there is an urgent need to develop \textit{quantum software engineering} (QSE) techniques to support quantum software applications in various domains throughout the quantum software life cycle~\cite{zhao2020quantum}. This cycle includes five different stages: requirements, design, implementation, testing, and maintenance. At each stage, software engineers need to employ a suitable quantum software technique to ensure the completeness of quantum software applications. For example, developers are required to model an architecture and understand the modularity of quantum software systems at the quantum software design stage. However, there are numerous quantum software techniques at each stage, posing challenges to correctly using these techniques. QSE provides guidelines to help developers select appropriate quantum techniques for fully developing quantum software applications. 

Software analytics is recognized as a critical part of developing classical software systems~\cite{menzies2013software, buse2012information, zhang2013software}. It aims to monitor, predict, and improve the efficiency and effectiveness of software applications during their implementation, testing, and maintenance stages. For example, Tasktop,\footnote{\url{https://www.tasktop.com/}} a software analytics tool, seeks to improve software quality by providing developers with a real-time view of how their software application is operating. As another example, Embold,\footnote{\url{https://embold.io/}} a software analytics platform, helps developers analyze their source code and improve its stability and maintainability. 
%\hoa{Give a few more examples} \jh{Anh @Hoa: Two examples, i.e., Tasktop and Embold, are enough?}

Similar to classical computing, quantum computing also needs software analytics to understand software artifacts, such as source code, bug reports, commits, etc., to assist developers in making better decisions in implementing quantum software applications.  As quantum computing employs the principles of quantum mechanics to process data, we need to improve traditional software analytics to better understand the quantum computing components, such as qubits, quantum logic gates and quantum algorithms. In this case, we can improve quantum software quality, accelerate productivity, and reduce quantum software maintenance costs. 

%Software analytics can also employ machine learning (ML) techniques to derive analysis results from software data~\cite{zhang2011software, dong2018feature, gousios2018big}. Similarly, quantum computing also requires these techniques for providing accurate judgments throughout the process of producing quantum software products. Specifically, ML methods can be employed to estimate the effort of developing quantum applications, migrate the current classical software code to quantum software code, generate quantum code from quantum specifications, and predict defects in quantum software applications. 

% - Machine learning in quantum computing

In this paper, we present the opportunities and challenges of software analytics in building quantum software applications. We believe that software analytics is vital to reducing quantum software development costs and improving quality and speed to market. We identify a number of areas that will be critical to the success of software analytics in developing quantum software applications. Those areas represent a new set of problems for the software analytics community to explore. We also present a brief roadmap of how those new problems could be addressed. 

%\noindent \textbf{RQ1:} What are the main research problems in building quantum software applications?

%\noindent \textbf{RQ2:} How do we employ software analytics to solve these problems?

%\noindent \textbf{RQ3:} How do we evaluate the performance of our software analytics models?

%Much research work has been deeply investigated into employing software analytics to develop classical software applications~\cite{menzies2013software, buse2012information, zhang2013software}. We believe that the research work should be useful to develop software analytics models that can be applicable in the process of developing quantum software applications, thereby reducing the cost of building these applications. To address these research questions, researchers should draw together existing work employing software analytics from classical computing systems, state-of-the-art machine learning techniques, and the existing domain knowledge in quantum computing systems. In the rest of the paper, we aim to answer the research questions in greater detail.

% - Vision:
%     - Road map for software analytics 
% \section{Research Problems}
% \label{sec:problems}

% This section first presents the background knowledge of a quantum system. We then briefly describe the main research problems for developing quantum software applications. 

\section{Background}
\label{sec:quantum}

\begin{figure}[t!]
  \centering
  \includegraphics[width=1.0\linewidth]{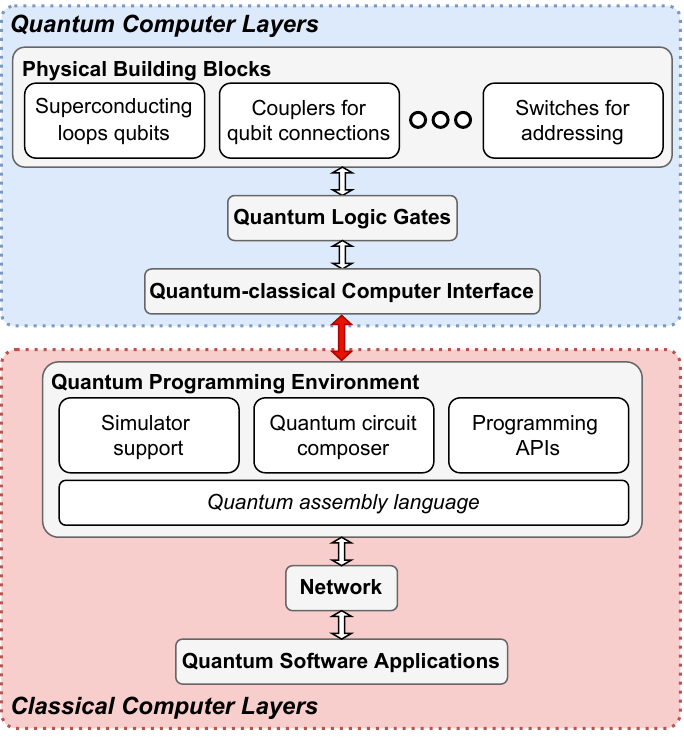}
  \caption{An overview of the architecture of a quantum computing system~\cite{sodhi2018quality}.}
  \label{fig:quantum}
\end{figure}

Quantum computing employs a quantum bit (qubit) to encode the information. Different from a classical bit, which has values of 0 and 1, each qubit $|e\rangle$ is represented by a linear combination of two basis states, such as  $|0\rangle$ and $|1\rangle$, in the quantum state space as follows:

% \begin{align*}
% |0 \rangle = 
%     \begin{bmatrix}
%       1 \\
%       0 \\
%     \end{bmatrix} \quad\quad\quad |1 \rangle = \begin{bmatrix}
%       0 \\
%       1 \\
%     \end{bmatrix}
% \end{align*}

\begin{align}
    |e\rangle = \alpha |0 \rangle + \beta |1 \rangle
\end{align}
where $\alpha$ and $\beta$ are the complex numbers in which $|\alpha|^{2}+|\beta|^{2}=1$. $|0\rangle$ and $|1\rangle$, the computational basis states of the qubit, are described as follows: 
\begin{align*}
|0 \rangle = 
    \begin{bmatrix}
      1 \\
      0 \\
    \end{bmatrix} \quad\quad\quad |1 \rangle = \begin{bmatrix}
      0 \\
      1 \\
    \end{bmatrix}
\end{align*}

% - What is quantum computing 

Figure~\ref{fig:quantum} shows an overview of the architecture of a quantum system~\cite{sodhi2018quality}. The architecture includes two main components: \textit{quantum computer layers} and \textit{classical computer layers}. The details of quantum computer layers are as follows: 

\begin{itemize} 
    \item \textit{Physical building blocks} have two vital parts: superconducting loops and couplers. While superconducting loops recognize the physical qubits, couplers connect different qubits in quantum systems. These blocks also contain other parts for qubit addressing and control operations. 
    \item \textit{Quantum logic gates}, the building blocks of quantum circuits, are used to process data in quantum systems.
    \item The \textit{quantum-classical computer interface} provides the interface between classical computers and a quantum processing unit (QPU).
\end{itemize}

The classical computer layers are described in the following: 

\begin{itemize} 
    \item The \textit{quantum programming environment} includes the quantum assembly language for instructing a QPU, the programming APIs used to write a high-level quantum programming language, and the simulator support employed to run and test quantum programs.
    \item A \textit{network} system connects the quantum programming environment and the quantum software applications.
    \item \textit{Quantum software applications}, written by developers, follow business requirements to serve customers. 
\end{itemize}

\section{Research Problems}
\label{sec:problem}

% \hoa{As discussed yesterday, we need to focus on articulating why these problems are novel. We need to pitch this paper like opening a new research area in software analytics/MSR. Make more use of Figure to explain the novelty and challenges of these new problems in quantum software analytics}

To build quantum software applications, we first need to estimate the cost of developing these applications. To simplify the quantum cost estimation problem, we neglect the cost of understanding customers' needs and designing the quantum system architecture. If stakeholders agree with the quantum applications' cost, developers will start writing a quantum program for the applications. During this process, developers need to deal with quantum software bugs that produce unexpected results. Specifically, a list of open main research problems in developing quantum software development are described as follows:

\noindent \textbf{1. Quantum software cost estimation:} Software cost estimation has been extensively investigated in the classical computing community~\cite{barry1981software, putnam1991measures, nassif2013towards, heiat2002comparison, huang2006optimization, kocaguneli2013kernel}. In quantum applications, stakeholders or software teams are also required to accurately predict the cost of these applications to ensure the success of their project.

Thus, there is a need for estimating for developing a quantum software application. This is especially important to decide the cost and benefit of developing a software application using quantum computing. Effort estimation for quantum applications represents a novel problem in software analytics due to their distinct characteristics. A quantum system is a hyprid system, including quantum computer layers and classical computer layers (see Figure~\ref{fig:quantum}). The \textit{physical building blocks} and the \textit{quantum programming environment} are the vital components of the quantum and computer layers, respectively. There are two main challenges in evaluating the effort of quantum software applications.

% At the early stage of developing quantum software systems, stakeholders or software project managers need to estimate the production cost of these systems. The success of quantum software projects mainly depends on the ability to accurately predict their development costs. Many research studies have been published to estimate the cost of building classical software systems~\cite{barry1981software, putnam1991measures, nassif2013towards, heiat2002comparison, huang2006optimization, kocaguneli2013kernel}. We classify these studies into two research approaches. The first research approach mainly focuses on constructing features, such as the number of lines of code, the number of developers involved in the software project, the complexity of the software project, etc., related to the software project cost~\cite{barry1981software, putnam1991measures}. The second research approach employs various machine learning techniques, such as neural networks~\cite{nassif2013towards}, analogy-based estimation~\cite{kocaguneli2013kernel}, regression trees~\cite{huang2006optimization}, etc., to accurately estimate the cost of building a classical software system. There are also studies combining the two approaches to improve the performance of classical software project cost estimation~\cite{mair2000investigation, finnie1997comparison}. 

\begin{itemize} [leftmargin=*]
    \item New models and techniques are needed for estimation the effort of constructing the \textit{quantum physical building blocks} in the quantum layers. As these building blocks include physical mechanisms such as \textit{superconducting loops} and \textit{couplers} (see Figure~\ref{fig:quantum}), developers require background knowledge in physics to correctly construct these building blocks. Research is needed to define a framework to estimate the knowledge of developers in comprehending the physical requirements of developing quantum software applications.
    
    \item We also need to evaluate how developers are familiar with the quantum programming environment (see Figure~\ref{fig:quantum}). During the development of quantum applications, developers are required to use suitable tools for simulating quantum computation (simulator support), optimizing quantum circuits (quantum circuit composer), describing quantum computation in a circuit model (quantum assembly language), and writing a quantum programming language (programming APIs). New research should investigate how developers comprehend these quantum tools to accurately estimate the effort for developing quantum applications.
\end{itemize}

% To accurately predict the cost of quantum software systems, there is a need to define a new list of features for estimating quantum systems' efforts. Specifically, we should consider two main things, i.e., the \textit{physical building blocks} and the \textit{quantum programming environment} (see Figure~\ref{fig:quantum}), when developing quantum software applications. It leads us to two following research questions:

% \begin{itemize}
%     \item How much time do developers need to spend on building quantum physical building blocks in a quantum software system? 
%     \item How are developers familiar with the quantum programming environment?
% \end{itemize}

\noindent \textbf{2. Quantum code migration:} Code migration is essential in the modern world of technology, where stakeholders often develop their products on multiple operating platforms using different programming languages~\cite{nguyen2015divide, nguyen2016mapping, emmerich2000implementing}. As quantum computing potentially outperforms classical computing in various domains, such as biochemistry, machine learning, or cybersecurity, many quantum programming languages, such as Qiskit~\cite{cross2018ibm}, ProjectQ~\cite{steiger2018projectq} and pyQuil~\cite{koch2019introduction}, have been developed. Therefore, there is a need to translate source code from classical programming languages to quantum programming languages to reduce the cost of implementing quantum software systems.

Code migration in quantum computing systems is a challenging task. The main reason is that it is difficult to understand quantum programming behavior. Unlike classical computing, where we can employ programming analysis techniques to analyze the behavior of classical programs, quantum computing uses qubits to encode information. We are clueless about how qubits are connected during the execution of a quantum program, leading to our incomprehension of the behavior of quantum programming. JavadiAbhari et al.~\cite{javadiabhari2015scaffcc} present an entanglement analysis that helps developers identify possible pairs of qubits to understand the behavior of quantum programs. However, it is unclear whether the analysis can be used on complex quantum programs.

\noindent \textbf{3. Quantum code generation:} Code generation is a vital problem in classical computing. Its goal is to generate explicit code from multimodel data sources, such as modeling languages~\cite{kundu2013automatic}, formal specification languages~\cite{darvas2016plc}, and natural language descriptions~\cite{xu2022ide}. Code generation is also a critical research problem in quantum computing to facilitate the process of developing quantum software applications. The main challenges of quantum code generation come from the data sources of quantum systems, such as quantum modeling languages and quantum specification languages. Unlike classical computing, where its modeling and specification languages have been deeply investigated, the research of quantum modeling languages and quantum specification languages has just started.

P{\'e}rez-Delgado and Perez-Gonzalez~\cite{perez2020towards} extended the unified model language (UML) to model quantum software systems. Their approach covers two types of UML, such as quantum class diagram and quantum sequence diagram. While the quantum class diagram indicates whether a software module makes use of quantum information, the quantum sequence diagram shows the connection between these software modules in a quantum program. However, P{\'e}rez-Delgado and Perez-Gonzalez have ignored diagrams for the vital components of quantum systems, such as superconducting loop qubits, quantum logic gates, or quantum circuit composers (see Figure~\ref{fig:quantum}). These components need to be further studied to construct a model language for the quantum system.

Cartiere~\cite{cartiere2016quantum} defined a formal specification language for quantum algorithms, but the language has only represented some elementary quantum logic gates, such as the Identity gate, C-Not gate, or Hadamard gate. In addition, the language has ignored the \textit{physical building blocks} (see Figure~\ref{fig:quantum}) of the quantum system. Even though the language can be used to specify a simple quantum system, its usefulness in complex quantum systems has been unknown.

Researchers need to investigate quantum modeling and specification languages to accurately solve the quantum code generation problem. Moreover, we need to develop a quantum verification program to ensure the generated code is consistent with the quantum system.

    % - Code generation from quantum spec (natural language, control language or formal language). 
    
\noindent \textbf{4. Quantum defect prediction:} Defect prediction is essential to support developers in releasing stable software applications~\cite{hall2011systematic, yang2015deep, hoang2019deepjit}. Defect prediction also plays an important role in reducing costs and improving the quality of quantum software systems. As quantum systems require a hybrid system, including quantum computer layers and classical computing layers (see Figure~\ref{fig:quantum}), many types of defects, such as incorrect quantum initial values, incorrect deallocation of qubits, and incorrect compositions of operations, have been found during the process of implementing quantum applications. There are two main challenges in detecting defects in quantum systems:

\begin{itemize} [leftmargin=*]
    \item Research in quantum software debugging and quantum software testing has received minor attention and still remains a vital problem in quantum systems~\cite{zhao2020quantum}. As the systems often have complex components, such as \textit{physical building blocks} and \textit{quantum logic gates} (see Figure~\ref{fig:quantum}), it can be challenging to find defects in their source code. Moreover, there is no prior work focusing on defining concrete defect patterns in quantum programming languages. 
    \item Developers require some knowledge of quantum computing systems to understand defects in their source code. However, it takes a lot of time, effort, and experience from developers during the process of developing quantum software applications. As quantum software applications have remained undeveloped, defects described by developers may not be correct in practice. 
\end{itemize}

\section{Initial Solutions}
\label{sec:sol_eval}

% \hoa{This section still reads like just applying existing approaches to a slightly new problem. I think we can reduce it to 0.5 pages and make it at a high level like a roadmap (don't need to discuss a solution for each problem, but present a high-level roadmap to the whole field. Use the space for Section 3.}

In this section, we present the solutions and an evaluation of the main research problems as follows:

\noindent \textbf{1. Quantum software cost estimation:} To evaluate the cost of quantum software systems, we should produce an effort estimation. Specifically, given a quantum software system $\mathcal{Q}$, the effort to implement the system is described as: 

\begin{equation}
    \mathcal{E}_{\mathcal{Q}} = \theta(f_1, \dots, f_n)
\end{equation}
where $\theta$ is the effort prediction function. $f_1, \dots, f_n$ is a list of features used to estimate the effort of implementing the quantum system. Specifically, the features are grouped into four different categories, such as product attributes, quantum system attributes, personnel attributes, and project attributes. The product attributes describe an overview of our product. The quantum system attributes, such as interoperability, security, or usability, focus on implementing the quantum system. The personnel attributes measure how familiar developers are with quantum systems. The project attributes present tools used in developing quantum systems.
% The product attributes describe an overview of our product, such as product complexity, product size, etc. The quantum system attributes focus on implementing the quantum system, such as quantum logic gates, quantum algorithms, hardware for quantum computing, etc. The personnel attributes measure how familiar developers are with quantum systems, such as quantum programming languages experience, quantum applications experience, etc. The project attributes present tools used in developing quantum systems, such as quantum programming languages, quantum software tools, etc. 

The cost estimation of quantum systems is then calculated by employing various methods, such as COCOMO~\cite{miyazaki1985cocomo}, Putnam~\cite{pillai1997model}, or function point-based analysis~\cite{low1990function}. For example, we can apply the Putnam method to define the cost estimation of a quantum system as follows: 

\begin{equation}
    \mathcal{C}_{\mathcal{Q}} = \mathcal{F}_{e} \times \mathcal{E}_{\mathcal{Q}}^{1 \mathbin{/} 3} \times t_d^{4 \mathbin{/} 3}
\end{equation}
where $t_d$ and $\mathcal{F}_{e}$ represent the delivery time of the quantum system and the competencies of quantum development, respectively. Both $t_d$ and $\mathcal{F}_{e}$ are taken by using past quantum system projects.  

\noindent \textbf{2. Quantum code migration:} As quantum computing potentially outperforms classical computing in terms of efficiency, many quantum programming languages have been developed for implementing quantum systems. Moreover, classical software systems have grown significantly nowadays, leading to a need to translate source code from classical programming languages to quantum programming languages. 

Researchers employ statistical machine translation techniques to solve the code migration problem in classical systems~\cite{nguyen2015divide, nguyen2016mapping, emmerich2000implementing}. We believe that these techniques are applicable in quantum code migration. Specifically, a classical code (a source code) is treated as a sequence of code tokens and is migrated into a fragment of a quantum code (a target code). In other words, we aim to map the classical code to the quantum code by analyzing the bilingual dual corpus, and then we extract the alignment between the tokens of the classical and quantum codes. We also need to manually define the translation rules for the mappings for the APIs used in the classical and quantum codes to improve the performance of our code migration models. For example, \texttt{sklearn.svm.SVR}\footnote{https://scikit-learn.org/stable/modules/generated/sklearn.svm.SVR.html} and \texttt{qiskit\_machine\_learning.algorithms.QSVR}\footnote{https://github.com/Qiskit/qiskit-machine-learning} are two APIs for calling a support vector regression model in Python (a classical programming language) and Qiskit (a quantum programming language), respectively. To estimate the performance of quantum code migration, we can employ the BLEU score as our evaluation metric~\cite{papineni2002bleu}. 
%Note that the BLEU score has been widely employed in code migration in classical computing~\cite{nguyen2015divide, nguyen2016mapping, emmerich2000implementing}.

\noindent \textbf{3. Quantum code generation:} Similar to code generation in classical computing, we can generate quantum code from various data sources, such as  quantum model languages, quantum specification languages, or natural language descriptions. However, the quantum model languages and the quantum specification require further study to employ them in developing quantum software systems in practice. 
% For example, Cartiere~\cite{cartiere2016quantum} defined a formal specification language for quantum algorithms, but the language has only been used to specify some simple quantum algorithms. The usefulness of this language for complex quantum algorithms has been unknown. P{\'e}rez-Delgado and Perez-Gonzalez~\cite{perez2020towards} extended the unified model language (UML) to model quantum software systems. However, some components, such as quantum activities, quantum state machines, etc., are obscure and need to be further studied to visualize the design of quantum systems. 

In classical computing, researchers often employ deep learning (DL) frameworks to generate code from natural language descriptions~\cite{xu2022ide}. These frameworks may be appropriate for generating quantum code to reduce the cost of developing quantum software applications. However, there are two main challenges to employing the DL techniques. First, this problem requires a large number of pairs of text descriptions and target quantum codes. For example, GitHub Copilot,\footnote{https://en.wikipedia.org/wiki/GitHub\_Copilot} an AI tool generating programming language codes from comments, trains a deep learning model from 54 million public Python GitHub repositories. As quantum code generation is a new research topic, it needs time for developers to build up the pairs of text descriptions and quantum codes. Second, different from classical computing, where its code structures are represented in various forms, such as abstract syntax trees, control flow graphs, or program dependency graphs, quantum code structures are still unexplored. These two challenges may lead to poor performance in implementing quantum code generation models. More research work needs to be done in the future to address the problem of quantum code generation.

\noindent \textbf{4. Quantum defect prediction:} Detecting defects in quantum systems is a critical research problem in developing any quantum software application. Like in classical computing, we can construct quantum defect prediction models based on high-quality quantum code metrics. The quantum code metrics should be related to the quantum system, such as:
\begin{itemize} [leftmargin=*]
    \item How many quantum logic gates are in the quantum system? What are they?
    \item How many quantum algorithms are employed in the quantum system? What are they? 
    \item What is the size of the quantum system? 
\end{itemize}
%Deep learning methods, such as attention networks~\cite{wang2017residual}, convolutional neural networks~\cite{gu2018recent}, or graph transformer network~\cite{yun2019graph}, 
Deep learning methods~\cite{wang2017residual, gu2018recent, yun2019graph}
% , such as attention networks~\cite{wang2017residual}, convolutional neural networks~\cite{gu2018recent}, or graph transformer network~\cite{yun2019graph}, 
can be employed to automatically extract high-quality code metrics for detecting defects in quantum systems. Another approach is to identify defect patterns that may happen in quantum programs. Zhao et al.~\cite{zhao2021identifying} show that there are some defect patterns in the quantum programming language Qiskit. We believe that pattern mining techniques~\cite{aggarwal2014frequent}, such as clustering or association rule learning, are appropriate to automatically identify such patterns to improve developers' productivity and reduce quantum software maintenance costs. Researchers can leverage a number of widely-used evaluation metrics, such as precision, recall, or F-measure, to capture the performance of their quantum defect prediction models.

\section{Conclusion}
\label{sec:conclusion}

% Quantum computing is powerful in terms of qubit counts, algorithms, and decoherence times. Stakeholders' interest in applying quantum computing has surged in recent years. Leveraging technology in solving scientific problems requires a deeper understanding of essential characteristics of quantum-specific applications, particularly those relevant to software development. As such, 
% more and more software applications can be facilitated by quantum computing, and the need for high-quality quantum applications will increase dramatically in the future. 
% % However, designing and implementing quantum computing applications is complex and requires different experts in various fields. 
% We believe that software engineering methodologies could be integrated into quantum computing applications. We propose a set of software analytics challenges and opportunities that likely affect quantum software applications.
% % As quantum computing is undergoing rapid development, we proposed a set of software analytics challenges and opportunities that likely affect quantum software applications.

Quantum computing is powerful in terms of qubit counts, algorithms, and decoherence times. Stakeholders' interest in applying quantum computing has surged in recent years. Leveraging technology to solve scientific problems requires a deeper understanding of the essential characteristics of quantum-specific applications, particularly those relevant to software development. As such, more and more software applications can be facilitated by quantum computing, and the need for high-quality quantum applications will increase dramatically in the future. We believe that software engineering methodologies need to be leveraged in quantum systems to help researchers and practitioners more easily construct quantum software applications.

\bibliographystyle{IEEEtran}
\bibliography{main}

\end{document}